\documentclass[11pt]{article}
\usepackage{amssymb, graphicx, hyperref}
\usepackage{amsmath}
\usepackage[margin=1in]{geometry}
\usepackage{float}
\usepackage{bm}
\usepackage[utf8]{inputenc}
\usepackage{xcolor}
\usepackage{caption}
\title{Colloidal hydrodynamic interactions in viscoelastic fluids}
\author{
    Dae Yeon Kim$^{1,*}$, Sachit G. Nagella$^{1,*}$, Saksham Malik$^{1}$,\\
    Nayeon Park$^{2}$, Jaewook Nam$^{2,3}$, Eric S.G. Shaqfeh$^{1,4}$, Sho C. Takatori$^{1,\dagger}$ \\
    \\
    $^{1}$Department of Chemical Engineering, Stanford University, CA, USA \\
    $^{2}$Department of Chemical and Biological Engineering, Seoul National University, Seoul, Korea \\
    $^{3}$Institute of Chemical Processes, Seoul National University, Seoul, Korea \\
    $^{4}$Department of Mechanical Engineering, Stanford University, CA, USA \\
    \\
    $^*$These authors contributed equally to this work. \\
    $^\dagger$Corresponding author: \texttt{stakatori@stanford.edu}
}

\date{\today}

\begin{document}
\maketitle

\begin{abstract}
The motion of suspended colloidal particles generates fluid disturbances in the surrounding medium that create interparticle interactions.
While such colloidal hydrodynamic interactions (HIs) have been extensively studied in viscous Newtonian media, comprehensive understanding of HIs in viscoelastic fluids is lacking.
We develop a framework to quantify HIs in viscoelastic fluids with exquisite spatiotemporal precision by trapping colloids and inducing translation-rotation hydrodynamic coupling.
Using solutions of wormlike micelles (WLMs) as a case study, we discover that HIs are strongly time-dependent and depend on the structural memory generated in the viscoelastic fluid, in contrast to ``instantaneous'' HIs in viscous Newtonian fluids.
We directly measure ``time-dependent'' HIs between a stationary probe and a driven particle during transient start-up, developing on the WLM relaxation timescale.
Following the sudden cessation of the driven particle, we observe an intriguing flow reversal in the opposing direction, lasting for a time $10 \times$ larger than the WLM relaxation time.
We corroborate our observations with analytical microhydrodynamic theory, direct numerical solutions of a continuum model, and particle-based Stokesian dynamics simulations.
We find that the structural recovery of the WLMs from a nonlinear strain can generate anisotropic and heterogeneous stresses that produce flow reversals and hydrodynamic attraction among colloids.
Measured heterogeneities indicate a breakdown of standard continuum models for constitutive relations when the size of colloids is comparable to the length scales of the polymeric constituents and their entanglement lengths.
\end{abstract}

\section{Introduction}

The dynamics of colloidal-scale particulates suspended in polymer-laden flows is important not only for fundamental understanding but also for numerous technological applications \cite{Shaqfeh2019-rn, Metzner1985-mf, Leal1979-cj}.
In zero-Reynolds-number Newtonian flow, the spatial decay of fluid disturbances is an algebraic function of the particle separation, notoriously giving rise to long-ranged hydrodynamic interactions (HIs) where the motion of immersed particles entrain one another in their disturbance flows.
The theory of HIs in viscous Newtonian media is well-established \cite{Stimson1926-nr, Batchelor1972-hq, Goldman1966-ik, Mazur1982-of, Kynch1959-jj, Happel2012-fp,  Kim1991-cs}.
For configurations involving several particles, the Stokesian dynamics method has been widely adopted to resolve the influence of many-body interactions \cite{Durlofsky1987-oc, Brady1988-ed, Brady1988-yd, Sierou2001-ks, Fiore2019-or}.
Newtonian HIs are ``instantaneous'' at zero-Reynolds-numbers, meaning that fluid disturbances propagated by moving colloids are established on negligibly fast timescales relative to particle motion, and therefore are solely dependent on a given colloidal configuration.
\begin{figure}[t!]
\centering
\includegraphics[width=15cm]{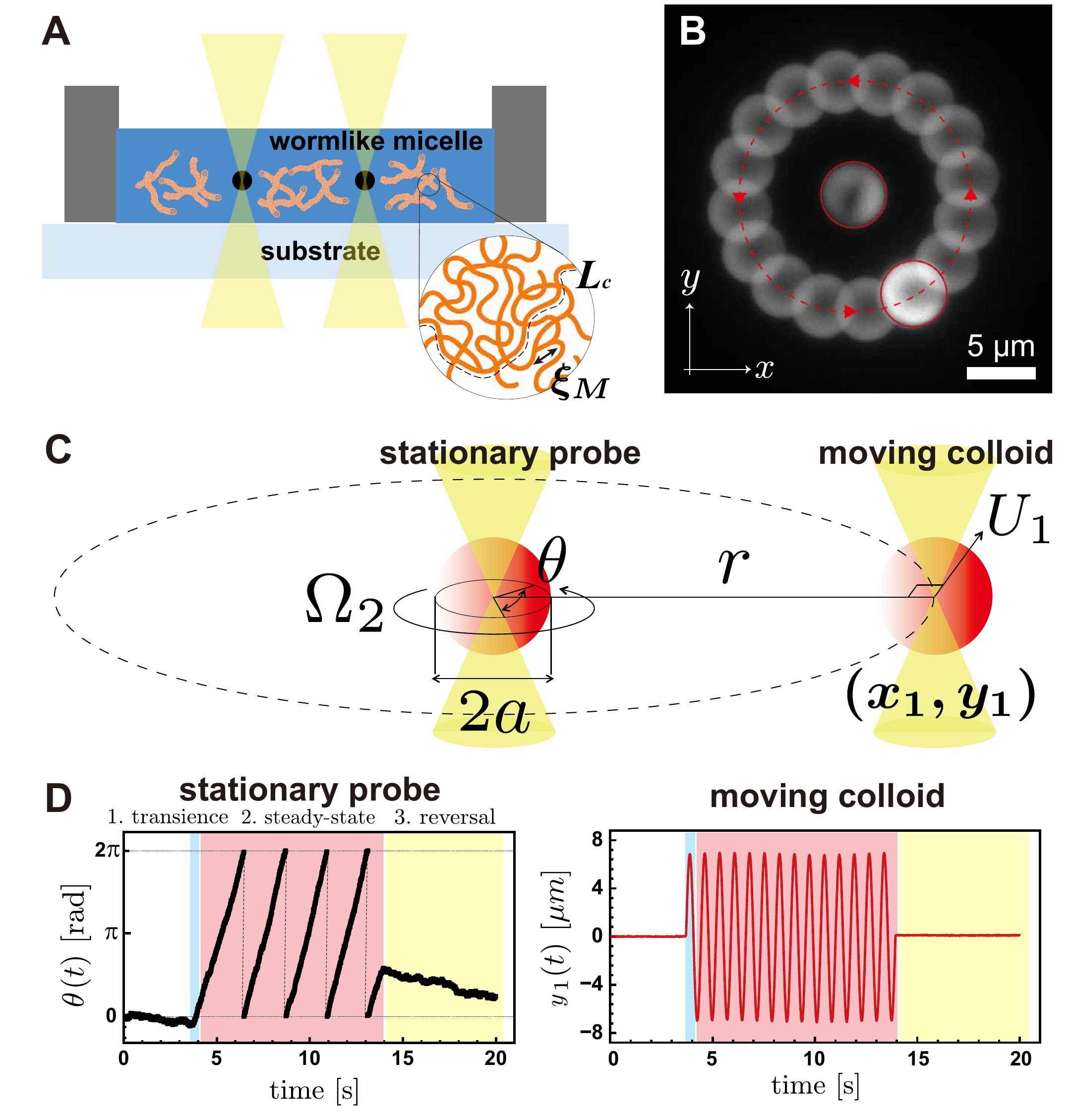}
\caption{Experimental setup of measuring the induced rotational motion a colloidal probe due to a nearby, translating particle in viscoelastic media. 
(A) Schematic of optical laser tweezers and trapped particles in wormlike micellar solution. 
(Close-up) Wormlike micelles self-assemble into an entangled network of flexible cylindrical chains with characteristic mesh size ($\xi_M \approx 120-200$ nm) and contour length ($L_c \approx 0.7-3.5$ \textmu m).(B) Superimposed image of many frames of a translating outer particle. See \textcolor{red}{Movie S1} for a video of the experiment. (C) We used strong traps to hold a pair of particles of radius $a = 2.5$ \textmu m at a fixed pair separation distance, $r$. Colloidal surfaces are labeled with a hemispherical fluorescent label to track the particle orientation. We imposed a circular trajectory on the moving particle with speed $U_1 = 60$ \textmu m/s. Measurements of induced particle rotations report fluid-mediated interactions between the particles. The probe rotates with velocity $\Omega_2 = a\skew{4}{\dot}{\theta}$, where $\theta$ is the measured angular displacement and $\skew{4}{\dot}{\theta}$ is the angular velocity.  (D) Left: sample time-series the probe's angular displacement, $\theta(t)$. Right: imposed trajectory of the moving particle along the y-direction, $y_1(t)$.  We identify three regimes of interest in the evolution of the induced angular displacements: (1) transient start-up, (2) steady-state development, and (3) relaxation after sudden secession of the driven colloid.}
\label{fig:Fig1}
\end{figure}
This is not necessarily true in viscoelastic suspending media.
Particle motion distorts the surrounding microstructure (e.g., polymer conformation), whose structural recovery depends on the relaxation of underlying constituents that are perturbed from their equilibrium configuration \cite{Leal1979-cj}.

Experimental measurement, analytical theory, and numerical simulation of colloidal dynamics in non-Newtonian media have each received significant attention.
Dilute suspension theories have been developed to model non-Newtonian flows around a single particle \cite{Frater1967-rj, apte2025average, neo2024stresslet, Einarsson2017-sb, zhang2020lift, einarsson2018einstein}.
Many experiments have focused on the behavior of an isolated probe \cite{Caspers2023-da, Ginot2022-lz, Gomez-Solano2015-yg, khan2019optical, ginot2022recoil}.
Two-body hydrodynamic coupling is critical for modeling the steady non-reciprocal motion of microswimmers in viscoelastic fluids \cite{Datt2018-xr, Joens2024-bp}.
In multiparticle systems, extensions to the Stokesian dynamics formalism have been developed \cite{Schaink2000-ob, Phillips1996-fy, Phillips2003-nj}.
For more general particle configurations and motion, direct numerical solutions to differential constitutive models compute non-Newtonian flow fields on a meshed Eulerian grid \cite{Yang2016-ax, richter2010simulations}.
Alternatively, viscoelastic flows around complex geometries have been modeled using mesh-free Lagrangian approaches, such as Lattice-Boltzmann \cite{Kuron2021-fx, Lee2017-mo} and smoothed particle hydrodynamics simulations \cite{Vazquez-Quesada2009-mf, Vazquez-Quesada2016-os, Vazquez-Quesada2017-dh, Hashemi2011-ao}.
Yet, many viscoelastic materials exhibit a broad spectrum of length and time scales, including micron-scale polymer entanglements and seconds-scale chain reptation times. 
These long lengths and slow time scales are comparable to those of the embedded colloids and their dynamics, so the colloids do not ``see'' a continuum medium with a known constitutive stress tensor.
There is currently no experimental method to systematically probe the many-body hydrodynamic interactions among colloids when the continuum approximation breaks down in the surrounding media.
To these ends, we present a combined effort of optical tweezer experiments, microhydrodynamic theory, direct numerical computation, and Stokesian dynamics simulations to study the fluid disturbances generated in a viscoelastic medium due to the motion of trapped colloids, using solutions of wormlike micelles (WLMs) as a case study.

Using optical tweezers, we focus a laser on a particle's center-of-mass and apply a stiff harmonic trap (Fig.~\ref{fig:Fig1}A).
We have shown recently \cite{kimnagella2025direct} that optical tweezers do not exert out-of-plane torques, enabling precise measurement of a tracer probe's angular displacement by tracking a hemispherical fluorescent label. 
In a two-particle setup, we drive one particle in circular orbits around a stationary probe (Fig.~\ref{fig:Fig1}B). 
The induced rotational velocity on the tracer probe provides a direct measure of disturbances in the suspending medium (Fig.~\ref{fig:Fig1}C).

We conduct experiments in solutions whose constituents impart viscoelasticity onto the bulk fluid, including short-chained polymers like polyethylene oxide (PEO) and long-chained, dynamic polymers like WLMs of cetylpyridinum chloride (CpyCl) and sodium salicylate (NaSal). 
WLMs can self-assemble into an entangled network of flexible cylindrical chains, resembling a semi-dilute polymer solution \cite{Lequeux1994-ef} (see close-up schematic in Fig.~\ref{fig:Fig1}A).
Linear viscoelasticity of WLM solutions is well-described by the reptation-reaction model of Cates \cite{cates1987reptation, Spenley1993-os}, in which stress relaxes over a single mode, $\tau(t) \sim \text{e}^{-t/\lambda}$.
The relaxation time, $\lambda$, is the geometric mean of chain reptation (slow) and micellar breakdown (fast).
At early times of the experiments, we observe transient start-up in the induced angular displacement of the probe, prior to steady development (see blue and red highlighted region of Fig.~\ref{fig:Fig1}D in plot titled ``stationary probe'').
We develop a theoretical microhydrodynamic framework to model the colloidal HIs that give rise to this transient behavior. We use our analytical model to measure the local relaxation time of WLM solutions, providing valuable information on the spatial heterogeneities in WLM media that are typically averaged out in traditional macrorheometry.

After driving the moving particle around the probe several times and stopping, we find that the probe's angular displacement can reverse for several seconds, which is dramatically longer than the milliseconds-scale relaxation times (see yellow highlighted region of Fig.~\ref{fig:Fig1}D).
While our goal is to develop a framework for studying HIs in general viscoelastic fluids, we focus on WLMs in this work to allow comparisons with the significant prior work on the rheology of WLMs.
The Gieskus model describes shear-thinning of WLMs at moderate shear-rates \cite{Kate-Gurnon2012-bg, Yesilata2006-iu}.
At higher shear-rates, shear-banding is observed \cite{Helgeson2009-fh, Helgeson2009-ir}.
Rolie-Poly models and their variations carefully consider the impact of micellar recombination kinetics on the fluid stress \cite{Salipante2024-hy, Salipante2024-td, Sato2022-nm}.
We are concerned with the role of elasticity of the entangled network on the observed reversal in the probe's angular displacement. For simplicity, we compare our experimental measurements to direct numerical solutions of the upper-convected Maxwell-Oldroyd model.
We obtain flow reversals that are much weaker and shorter-lasting, revealing a breakdown of the continuum model on colloidal scales.
To gain further mechanistic insight, we adopt a bead-spring representation of a heterogeneous elastic network, 
where individual beads are ``nodes'' that model hooks and junctions \cite{Vaccaro2000-vr, Vasquez2007-rp, Hernandez-Cifre2003-ru}.
We perform Stokesian dynamics simulations that couple the elastic forces exerted by bead-spring chains onto the rotational motion of trapped colloids. 
The results of our simulations suggest that WLM chain-chain entanglements spanning colloidal lengths are critical for producing the observed nonlinear creep response of our tracer probe.

\section{Materials and Methods} \label{sec:mat_methods}

\subsection{Wormlike micelle solutions}

We used aqueous surfactant solution composed of cetylpyridinium chloride (CPyCl, MP Biomedicals) and sodium salicylate (NaSal, Sigma-Aldrich) dissolved in Milli-Q water. The molar ratio of NaSal to CPyCl was fixed at 0.6. A stock solution containing 100 mM CPyCl and 60 mM NaSal was prepared and subsequently diluted with either Milli-Q water or 0.5\,M NaCl solution to create three test samples: 0.25\,wt\% in water, 0.25\,wt\% in 0.5\,M NaCl, and 0.125\,wt\% in 0.5\,M NaCl. Consequently, the final concentrations of the surfactant and counterion were 25\,mM and 15\,mM, respectively, for the 0.25\,wt\% samples, and 12.5\,mM and 7.5\,mM for the 0.125\,wt\% samples. All solutions were equilibrated at room temperature for 2 days prior to use.
Details on mechanical characterization are in the \textcolor{red}{SI Appendix}.

\subsection{Colloidal particles with hemispherical fluorescent coating}

Fluorescently labeled poly(methyl methacrylate) (PMMA) particles (5\% w/v, radius 2.5 \textmu m, Abvigen Inc.) were prepared by tagging amino-functionalized PMMA colloids with Alexa Fluor 647 (Invitrogen, USA) on the particle hemispheres, using a gel trapping technique \cite{paunov2003novel}. See \textcolor{red}{SI Appendix} for more details.

\subsection{Optical trapping and imaging setup}

Colloidal particles were manipulated using a commercial optical tweezers system (Tweez 305, Aresis Ltd.) equipped with a 1064\,nm infrared laser (maximum power 5\,W). Imaging was performed using an inverted fluorescence microscope (Ti2-Eclipse, Nikon) with a 100$\times$ oil immersion objective lens (NA = 1.45) and a sCMOS camera (Kinetix22, Teledyne). Fluorescence excitation was provided by a SpectraX LED light source (Lumencor) at 647\,nm, and emission was filtered using a 680/42 bandpass filter (Semrock, IDEX).
Additional details are in our previous work \cite{kimnagella2025direct} and \textcolor{red}{SI Appendix}. All experiments were performed at sufficiently large distances from the wall to minimize confinement effects.

\subsection{Measurement of angular displacement}

The angular displacement, \( \theta(t) \), of the stationary probe was measured by tracking the brightest point of the hemispherical fluorescent label. 
We used an in-house MATLAB script to extract the angular dependence of the fluorescence intensity profile along the particle perimeter, for each frame. 
The profile was fit to a Fourier series, and the angular position of maximum intensity was assigned as the particle orientation.

\subsection{Numerical methods}

We corroborated our experimental observations with direct numerical solutions to the upper-convected Maxwell-Oldroyd model, which were evaluated using computing resources provided by Anvil \cite{Song2022-gl}.
We also performed Stokesian dynamics simulations of trapped colloids interactions with bead-spring chains through a Newtonian fluid background. 
See \textcolor{red}{SI Appendix} for more details.

\section{Results}\label{sec:results}
\subsection{The Green's function for Stokes flow in a Maxwell fluid}
We study the fluid-mediated disturbances that are generated between trapped colloidal particles in viscoelastic solutions. 
To guide our microhydrodynamic analysis, we first develop the fundamental solution to inertialess flow in a Maxwell fluid.
A time-dependent point force acts on an unbounded fluid medium, and we wish to determine the resulting velocity disturbance, $\bm{u}(\bm{x},t)$.
The fluid stress is $\bm{\sigma}(\bm{x},t) = -p\mathbf{I} + \bm{\tau}$, where $p(\bm{x},t)$ is the pressure. 
We have the following coupled balance laws for momentum, mass, and viscoelastic stress,
\begin{align}
    \label{eq:cauchy}
    \nabla \cdot \bm{\sigma} + &\bm{F}(t)\delta(\bm{x}) = \bm{0}, \\
    \label{eq:continuity}
    \nabla &\cdot \bm{u} = 0,  \\ 
    \lambda \frac{\partial}{\partial t}\bm{\tau} + \bm{\tau} &+ \mathcal{O}(\text{Wi})= 
    2\eta_0\skew{3}{\dot}{\bm{\gamma}}
    \label{eq:stress_bal},
\end{align}
where $\eta_0$ is the zero-shear viscosity and $\skew{3}{\dot}{\bm{\gamma}}(\bm{x},t) = \frac{1}{2}\left( \nabla \bm{u} + \nabla \bm{u}^{T} \right)$.
We neglect the nonlinear couplings between flow and stress in the upper-convected Maxwell-Oldroyd model that scale with the Weissenberg number, $\text{Wi} = \skew{3}{\dot}{\gamma}_0 \lambda$. 
The characteristic shear-rate is $\skew{3}{\dot}{\gamma}_0$.

We define the Fourier transform pair, $\hat{f}(\omega) = \int_{-\infty}^{\infty} f(t)\text{e}^{-i\omega t} \text{d}t$ and $f(t) = \int_{-\infty}^{\infty} \hat{f}(\omega)\text{e}^{i\omega t} \text{d}\omega$.
Next, we express Eqn.~\ref{eq:stress_bal} in the frequency domain, 
\begin{equation}
    \hat{\bm{\tau}}(\bm{x},\omega) = 2\eta_0 \left( \frac{1}{1-i\omega\lambda} \right)\skew{3}{\hat}{\skew{3}{\dot}{\bm{\gamma}}}(\bm{x},\omega), 
\end{equation}
thereby obtaining a Newtonian-like constitutive relation that we insert to the Fourier transform of  Eqn.~\ref{eq:cauchy}. 
We then have

\begin{align} \label{vel_field_eqn}
 \eta_0\left( \frac{1}{1-i\omega\lambda} \right)\nabla^2\hat{\bm{u}} - &\nabla\hat{p} + \hat{\bm{F}}(\omega)\delta(\bm{x}) = \bm{0} , \\ 
 \nabla &\cdot \hat{\bm{u}} = 0 .
\end{align}

The velocity field in the frequency domain obeys the equations of Stokes flow, now with a complex-valued viscosity. The fundamental solution is therefore

\begin{equation} \label{eq:vel_field_freq}
 \hat{\bm{u}}(\bm{x}, \omega) = \frac{1}{8\pi \eta_0}\left( 1-i\omega\lambda \right) \mathbf{J}(\bm{x}) \cdot \hat{\bm{F}}(\omega),
\end{equation}
in which $\mathbf{J}(\bm{x}) = \mathbf{I}/r + \bm{x}\bm{x}/r^3 $ is the Oseen tensor, and $r = \left(\bm{x} \cdot \bm{x}\right)^{1/2}$. 
We will start from Eqn.~\ref{eq:vel_field_freq} in subsequent analysis. 
For completeness, Eqn.~\ref{eq:vel_field_freq} in the time-domain is $\bm{u}(\bm{x},t) = \frac{1}{8\pi \eta_0}\mathbf{J}(\bm{x}) \cdot \left[ 1 + \lambda\frac{\text{d}}{\text{d}t}\right]\bm{F}(t)$.
Setting $\lambda=0$ yields the ``quasi-static'' solution to Newtonian Stokes flow.
\begin{figure}[t!]
\centering
\includegraphics[width=15.8cm]{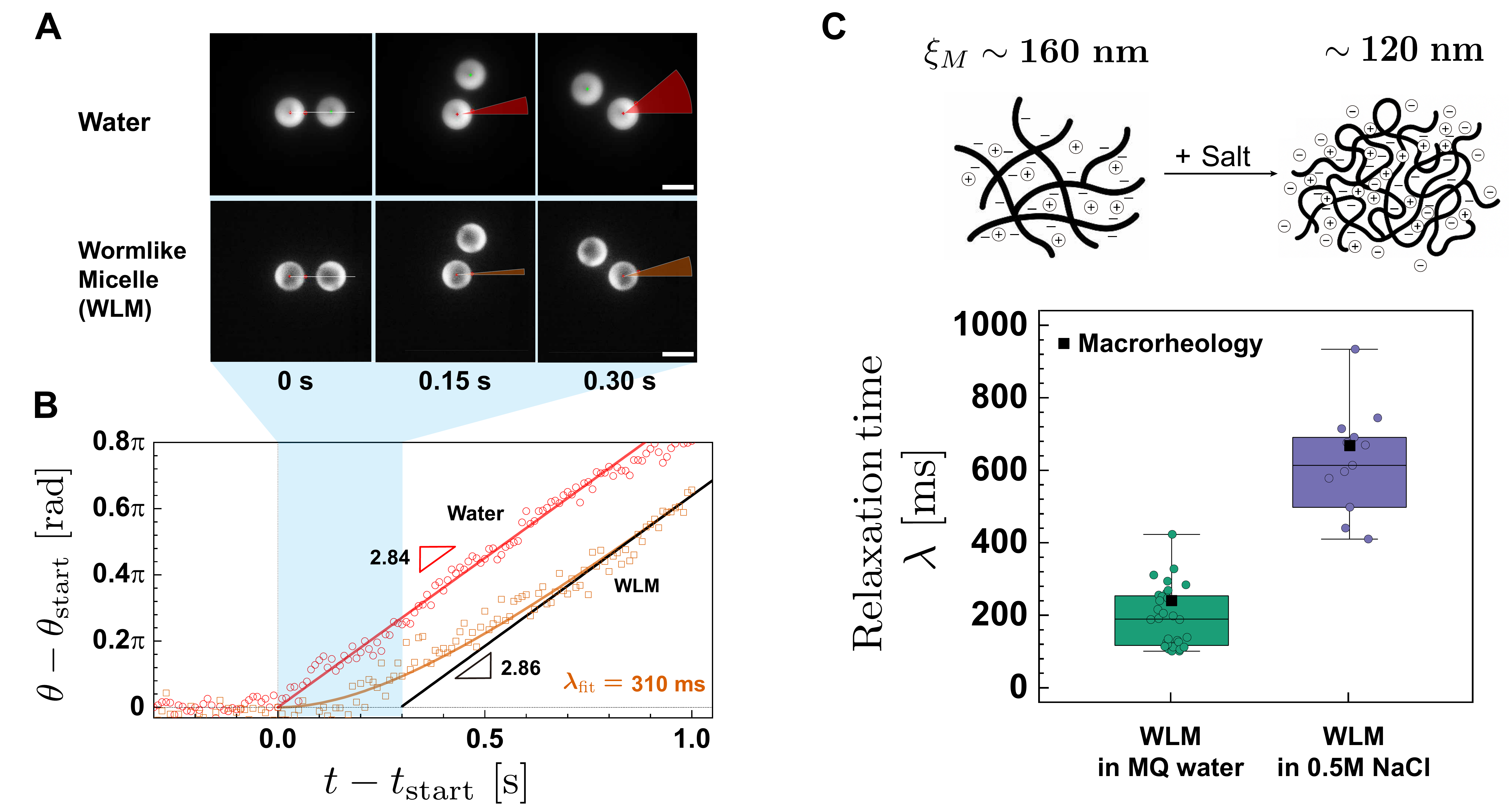}
\caption{
Fluid disturbances in viscoelastic media exhibit a delayed response to imposed driving forces. 
For a pair of spheres with radii $a=2.5$ \textmu m, we held the position of a probe particle and imposed a circular trajectory on the other with a specified radius to maintain a specified center-to-center distance (see Fig.~\ref{fig:Fig1}). Solvent-mediated forces generate a torque that rotates the probe. (A) Angular displacement of the probe in water and in a wormlike micellar (WLM) solution. Shortly after the driven particle starts moving, the swept angle is smaller in the WLM solution. Shaded regions highlight the induced angular displacement over the same time interval. Scale bar is 5 \textmu m. See \textcolor{red}{Movie S2} for a video of the experiment. (B) At a pair separation $r = 2.8a$ between the two particles, we measured the angular displacement of the probe as a function of time, starting from when the outer particle begins moving ($t-t_{\text{start}} = 0$) in a Newtonian solvent (red points) and in WLM solution (orange points). The translational velocity of the driven particle was 60 \textmu m/s.
In these sample data, the micellar solution is 0.125 wt\% in 0.5M NaCl. We fitted the transient start-up to our model for induced angular displacement of the probe, Eqn.~\ref{eq:ang_disp}, to obtain the relaxation time, $\lambda$, of the WLM solution (orange curve). 
The probe then assumes a steady-state rotational velocity (black line). 
For comparison, we performed the same measurement in water (red line). 
(C) Measurements of relaxation times in two different WLM solutions, as measured using our method in panel (B) (box plots). We obtain a distribution of relaxation times, upon varying the placement of the two spheres. The median values of the relaxation times are 189 ms in MQ water and 614 ms with added salt. Corresponding control measurements of the crossover point from frequency sweeps using a conventional macroscopic rotational rheometer are 250 ms and 670 ms (black squares). See Fig.~\textcolor{blue}{S2} in \textcolor{red}{SI Appendix} for an expanded version of this figure that includes data for different WLM concentrations.
}
\label{fig:Fig2}
\end{figure}

\subsection{Transient start-up} 
During the initial start-up of the driven particle, we observe that the angular displacement of the stationary probe is smaller in the WLM solution than that in water (compare the area of the highlighted regions between experiments conducted in water and in WLM solution in Fig.~\ref{fig:Fig2}A).
In water, the probe attains its steady-state angular velocity instantaneously (see red circles and solid line in Fig.~\ref{fig:Fig2}B).
In the WLM solution, however, the angular velocity develops over the relaxation timescale of the fluid ($\lambda = 303 \, \text{ms}$), before reaching a steady-state that is equivalent to that of water (see orange points and black line in Fig.~\ref{fig:Fig2}B). 
Therefore, the fluid disturbance generated by the driven particle in the WLM solution exhibits a delayed response to the imposed motion.

The fact that the spatial dependence in Eqn.~\ref{eq:vel_field_freq} is the same as in Newtonian media enables us to adopt a microhydrodynamic framework to describe colloidal HIs in a linear viscoelsatic medium. 
From Eqn.~\ref{eq:vel_field_freq}, the coupling between the particle kinematics and their corresponding multipole moments, truncating at the hydrodynamic torques, is \cite{kimnagella2025direct, Fiore2018-ax, Fiore2019-or} 
\begin{equation} \label{eq:two_body_mm}
    \begin{pmatrix}
        \hat{\textbf{U}} \\
        \hat{\bm{\Omega}}
    \end{pmatrix} \hat\varphi^{-1}(\omega)
    = 
    \begin{pmatrix}
        \textbf{M}^{\text{UF}} & \textbf{M}^{\text{UL}} \\
        \textbf{M}^{\Omega \text{F}} & \textbf{M}^{\Omega \text{L}}
    \end{pmatrix}
    \begin{pmatrix}
        \hat{\mathbf{F}}^{\text{H}}
        \\
        \skew{-4}{\hat}{\mathbf{L}}^{\text{H}}
    \end{pmatrix},
\end{equation}
where $\hat{\varphi}(\omega) = (1 - i\omega\lambda)$.
We use stacked vectors and block matrices for brevity. 
For example,  $\hat{\mathbf{U}}(\omega) = \left[\hat{\bm{U}}_1(\omega) \quad \hat{\bm{U}}_2(\omega)\right]^{T}$.
We can write Eqn.~\ref{eq:two_body_mm} succinctly as the frequency-dependent mobility relation, $\mathcal{M}(\bm{r}^N) \cdot \hat{\mathcal{F}}(\omega) = \hat{\mathcal{U}}(\omega)\hat{\varphi}^{-1}(\omega)$. 
$\mathcal{M}(\bm{r}^N)$ is the configuration-dependent grand mobility matrix \cite{Brady1988-ed, Fiore2019-or} that stores all pair hydrodynamic interactions.
For freely rotating particles, $\skew{-4}{\hat}{\mathbf{L}}^{\text{H}} = \bm{0}$.

We solve for the required hydrodynamic forces to maintain the imposed translational motion, and use these to determine the rotational velocities.
The result, after applying the convolution theorem, is
\begin{equation} \label{eq:tr_time_convolution}
        \int_{-\infty}^{t} \bm{\Omega}(s)\varphi^{-1}(t-s)\text{d}s = \\ 
        \textbf{M}^{\Omega \text{F}} \cdot \left( \textbf{M}^{\text{UF}} \right)^{-1} \cdot \int_{-\infty}^{t} \textbf{U}(s)\varphi^{-1}(t-s) \text{d}s.
\end{equation}
Equation \ref{eq:tr_time_convolution} shows that the rotational velocities are determined by the convolution of the imposed linear velocities with the Maxwell response kernel, $\varphi^{-1}(t) = \text{e}^{-t/\lambda}$.

The circular motion of the driven particle requires that its velocity is perpendicular to the vector of separation from the stationary probe.
Without loss of generality, we suppose that $\bm{U}_1(t) = U_1 \Theta(t)\bm{e}_{y}$ and $\bm{r}_{21}/r = \bm{e}_{x}$, so that $\bm{\Omega}_2(t) = \Omega_2(t)\bm{e}_z$.
$\Theta(t)$ is the Heaviside step function. 
The rotational velocity of the probe during the transient start-up is 
\begin{equation}\label{eq:tr_time_HI}
    \begin{split}
        \int_{-\infty}^{t} \Omega_2(s)\varphi^{-1}(t-s)\text{d}s =
        y^{*}(r/a)(1-\text{e}^{-t/\lambda})U_1,
    \end{split}
\end{equation}
where the pair hydrodynamic function is \cite{kimnagella2025direct}

\begin{equation} \label{eq:tr_HI_ss}
    y^{*}(x) = \frac{3x}{-4 -6x^2 + 8x^3} + \frac{3x}{4 + 6x^2 + 8x^3} .
\end{equation} 
At a given center-to-center distance, we integrate Eqn.~\ref{eq:tr_time_HI} and obtain the probe's angular displacement, 
\begin{equation} \label{eq:ang_disp}
    \theta(t; r) = \theta_{\text{start}} + y^{*}(r/a)\left[ t - \lambda \left( 1 - \text{e}^{-t/\lambda} \right) \right] .
\end{equation}

\subsubsection{Measurement of relaxation time}
We can measure the relaxation time by fitting Eqn.~\ref{eq:ang_disp} to the transient start-up in the angular displacement of the probe (see solid orange line in Fig.~\ref{fig:Fig2}B).
We performed the experiments in two different WLM solutions, which are prepared by dissolving CPyCl (with NaSal) in either Milli-Q (MQ) water or 0.5\,M NaCl solution, to measure their relaxation times (Fig.~\ref{fig:Fig2}C).
To validate our microscopic measurements, we performed frequency sweeps in a rotational rheometer to probe the elastic and viscous moduli within the linear viscoelastic regime.
From their crossover frequencies, the relaxation times were 250 and 670 ms, respectively (see black squares in Fig.~\ref{fig:Fig2}C, and see Fig.~\textcolor{blue}{S1} in \textcolor{red}{SI Appendix} for rheometric data). 
Using optial tweezers, we obtained a distribution of relaxation times, upon varying the placement of the two spheres in the viewing chamber (compare green and purple box plots with points in Fig.~\ref{fig:Fig2}C). 
The median values of the relaxation times were 189 and 614 ms, respectively, agreeing with those obtained from macrorheology. 

In the WLM solution, the disturbance generated by the driven particle propagates through an entangled network of polymeric chains. 
Although the shear rates in the interparticle gap are comparable to the WLM relaxation time, $\text{Wi} \sim \mathcal{O}(1)$, the agreement between our linear response analysis and the observed transience suggests that the nonlinear coupling of local shear onto stretch of the elastic network is negligible at early times.

The timescale over which hydrodynamic forces are transmitted to the tracer probe is set by the dynamics of polymeric chains.
Further, adding salt screens charges, promoting entanglements that thereby shrink the characteristic mesh size of the network \cite{rehage1988rheological}.
Between the two WLM solutions, with and without NaCl, we estimate the mesh sizes to be $\xi_M =$ 160 nm and 120 nm, respectively (See \textcolor{red}{SI Appendix} for calculation details).
A given chain takes longer to reptate in a denser network, ultimately slowing down the response of the WLM medium to the hydrodynamic forces generated by the driven colloid (compare estimated mesh sizes and corresponding relaxation times in Fig.~\ref{fig:Fig2}C).
We attribute the observed variation of relaxation times to spatial heterogeneity in entanglement density. Therefore, colloidal particles probe local structural heterogeneities in the WLM medium on mesoscopic length scales that are inaccessible using conventional macrorheometry.

\begin{figure}[t!]
\centering
\includegraphics[width=14cm]{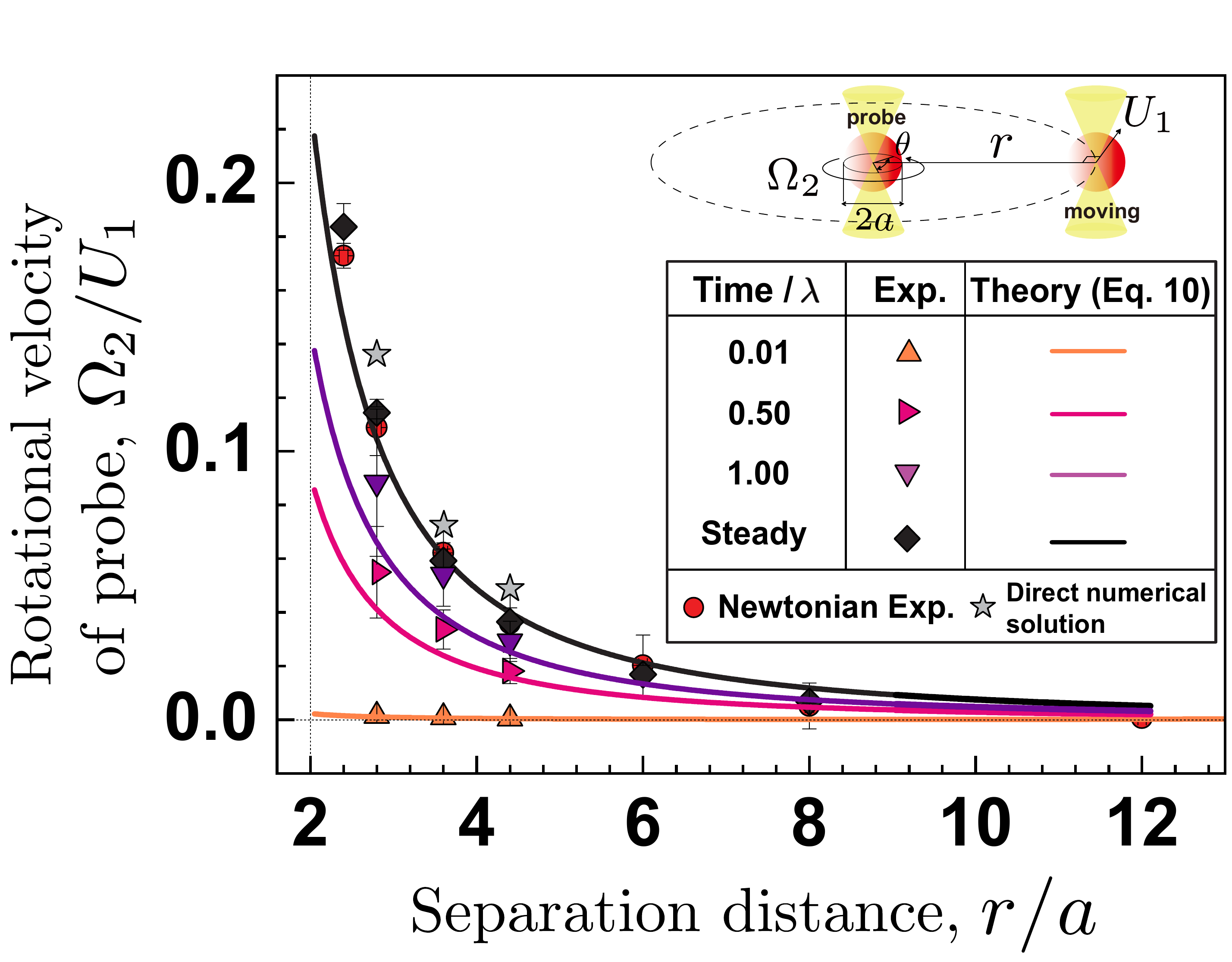}
\caption{
Wormlike micelle solutions exhibit a time-dependent hydrodynamic function during transient start-up.
For a pair of spherical particles of radius $a=2.5$ \textmu m, imposing a circular trajectory on one of the spheres generates fluid-mediated forces that rotate a probe with rotational velocity $\Omega_2$. 
We measured the rotational velocity of the probe at specified times: $0.01\lambda$,  $0.5\lambda$, and $1.0\lambda$ (colored triangles). After transient start-up, the probe assumed a steady rotational velocity (black diamonds).  For comparison, we plot measurements of the pair hydrodynamic interaction in water (red points). Solid curves are corresponding theoretical predictions using Eqn.~\ref{eq:tr_time_HI}.
Aside from the relaxation time, $\lambda$, fitted to the angular displacement data in Fig.~\ref{fig:Fig2}, there are no other adjustable parameters in our theory. 
Our steady-state data are compared against direct numerical solutions to the upper-convected Maxwell-Oldroyd model (gray stars).
See \textcolor{red}{SI Appendix} for further computational details.
}
\label{fig:Fig3}
\end{figure}

\subsubsection{Time-dependent pair hydrodynamic interaction}
We repeat our measurements at increasing center-to-center distances to measure a spatiotemporal hydrodynamic function between the two particles during the transient start-up (see filled symbols in Fig.~\ref{fig:Fig3}). 
Each point in Fig.~\ref{fig:Fig3} represents an average of the tracer probe's induced rotational velocity across different WLM solutions of varying surfactant and NaCl concentrations. 
Correspondingly, we present theoretical predictions of the transient, pair hydrodynamic function using Eqn.~\ref{eq:tr_time_HI} (see solid curves). 
Aside from the relaxation time, $\lambda$, fitted to the angular displacement data in Fig.~\ref{fig:Fig2}, there are no other adjustable parameters in our theory. 

At a given separation distance, the disturbance flow entraining the probe develops on the relaxation time of the linear Maxwell medium (compare data at times $t/\lambda = 0.01 -1.0$). 
At steady-state, the measured pair hydrodynamic function is equal to that in water (compare black diamonds to red circles).

These are compared against direct numerical solutions of Eqn.~\ref{eq:cauchy} and Eqn.~\ref{eq:continuity} using the upper-convected Maxwell-Oldroyd model instead of Eqn.~\ref{eq:stress_bal} (compare black data to gray stars). See \textcolor{red}{SI Appendix} for detail on the direct numerical solutions. Overall, we obtain excellent agreement.

So far, we have only discussed the first $\sim0.3-0.6$\,s of our experiments, just after the driven particle begins to move.
At these early times, highly entangled WLM networks are weakly perturbed from their equilibrium configuration, generating fluid stresses that are well-described using linear viscoelastic theory.
However, our model no longer holds after driven colloid strains the suspending medium for $\sim 10$\,s and significantly deforms WLM networks.
As a result, the tracer probe exhibits a nonlinear creep response that we discuss in the following section. 

\subsection{Reversal in angular displacement}
Now, we concern ourselves with the nonlinear regime of viscoelasticity, after the moving particle completes several orbits and strains the suspending WLM medium far away from linear perturbations.
After several orbits, we immediately stop the moving colloid to observe the creep response of the stationary probe's angular displacement. 
Interestingly, we observed a flow reversal that spun the stationary probe in the opposite direction.
This flow reversal continued more than several seconds, a time $> 10 \times$ larger than the WLM relaxation time ($\lambda \approx 0.3\,\text{s}$). 

\begin{figure}[htbp]
    \centering
    \includegraphics[width=15cm]{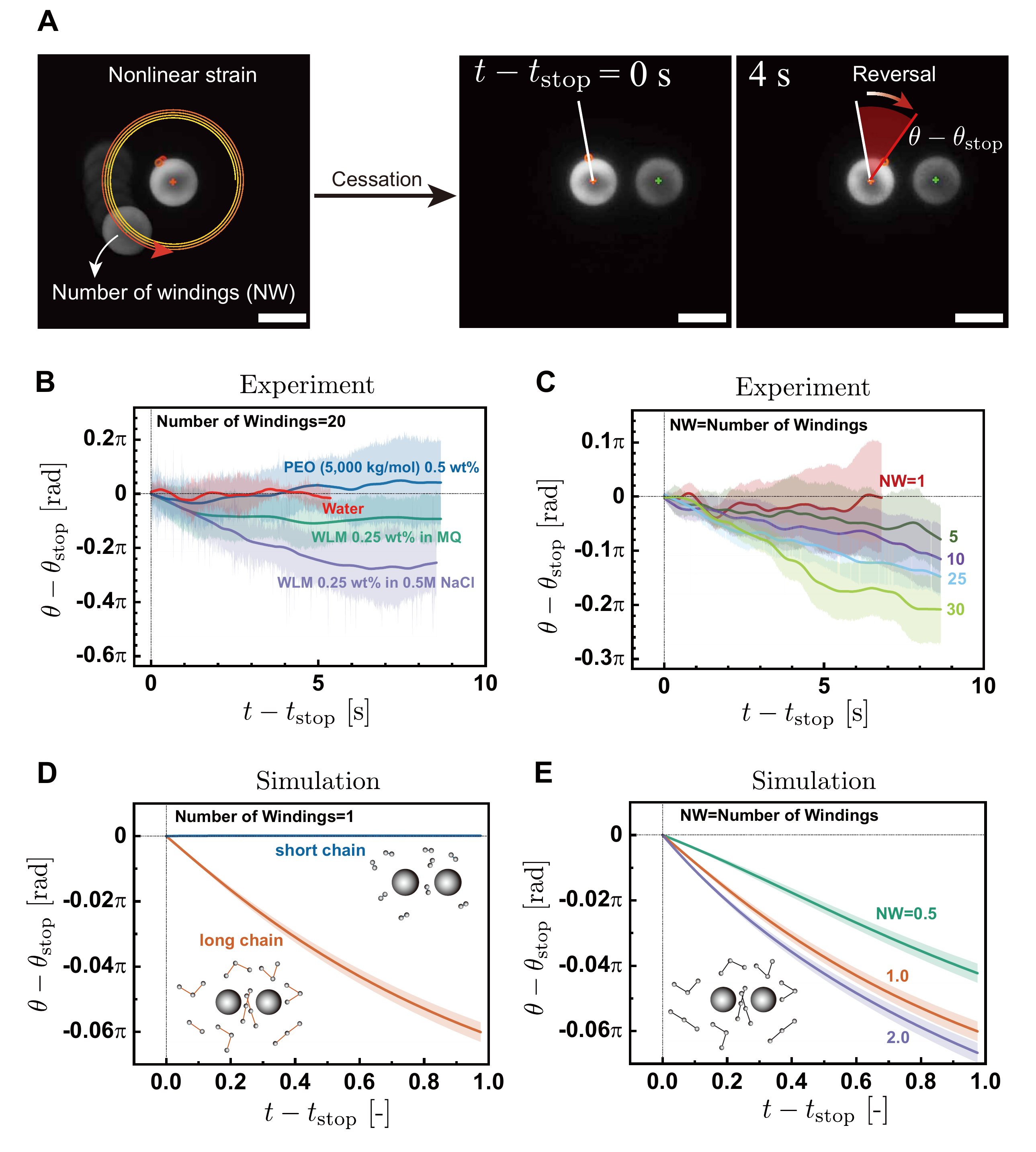}
    \caption{Flow reversal is observed after straining solutions of wormlike micelles (WLM).
    For a pair of spheres of radii $a = 2.5$ \textmu m, we measured the induced angular displacement, $\theta(t)$, of a fixed probe while the other was driven along a circular trajectory whose radius was maintained at a center-to-center distance $r = 2.8a = 7$\,\textmu m. (A) Reversal in angular displacement of the probe after stopping the driven sphere (``cessation'') at the time $t = t_{\text{stop}}$. Snapshots show the probe’s reversal in angular displacement after cessation, following several orbits made by the driven particle (``windings''). The scale bar represents 5~\textmu m. See \textcolor{red}{Movie S3} for a video of the experiment.
    (B) Measurement of reversal in various suspending media after cessation: 0.25 wt\% in MQ water and in 0.5\,M NaCl (see green and purple colors, respectively), polyethylene oxide (PEO) solution (blue data), and water (red data).
    (C) Dependence of reversal on the number of windings a WLM solution (0.25 wt\% in MQ water). A minimum of 8 and up to 16 repeat measurements were performed for each experimental condition. The solid line indicates the average response, while the shaded area represents the standard deviation across trials.
    (D) To corroborate these observations, we separately performed non-Brownian Stokesian dynamics simulations of two large spheres of radii $a_c$ in a bath of bead-spring chains with bead size $a_m = 0.1a_c$, modeling the behavior of colloidal particles in a viscoelastic medium.
    In all computations, the center-to-center distance was maintained at $r = 2.8a_c$.
    In the simulations with trimers, the angular displacement of the probe reverses after stopping the outer sphere (orange data). See \textcolor{red}{Movie S4} for a sample video of the simulation.
    We compare these results to simulations of the two spheres in a bath of dumbbells with shorter resting bond lengths (blue data). 
    In these simulation data, time is scaled by the moving sphere's period of orbit.  
    (E) Results of the angular displacement reversal from Stokesian dynamics calculations after winding the polymer bath 0.5 (green), 1.0 (orange), and 2.0 times (purple).
    Variation in simulation results arises from randomized initialization of bead-spring configurations.
    }
    \label{fig:Fig4}
\end{figure}

We systematically studied the dependence of this reversal on the number of times the driven colloid completed an orbit, which we define the ``number of windings'' (Fig.~\ref{fig:Fig4}A). 
We maintained the center-to-center distance at $7.0$ \textmu m. 

Results of the time-dependent reversal after 20 windings are shown in Fig.~\ref{fig:Fig4}B in various suspending media. 
In water, the angular displacement remains stationary after stopping the moving particle (see red data in Fig.~\ref{fig:Fig4}B).
In comparison, we observe reversal in WLM solutions dissolved in MQ water (see green data in Fig.~\ref{fig:Fig4}B).
Addition of a secondary salt, like NaCl, screens charges on the surfactant and counterion, assembling into longer, more flexible chains \cite{varade2005effect, haward2012stagnation}.
We performed our experiments in WLM solution with 0.5\,M NaCl, where the tracer probe exhibited greater amount of reversal (compare green and purple data in Fig.~\ref{fig:Fig4}B). 
We estimate the contour length of the WLM to be 3.5 and 0.72 \textmu m with and without NaCl, respectively (see \textcolor{red}{SI Appendix}).
The tracer probe exhibited greater reversal in the WLM solution with added NaCl (compare green and purple data in Fig.~\ref{fig:Fig4}B).
To highlight, 8 seconds after cessation, the amount of reversal was $0.28\pi$ and $0.04\pi$ with and without NaCl, respectively.

We next performed our experiments in 5,000 kg/mol polyethylene oxide (PEO) solution, whose chain sizes are much shorter than that of WLMs. 
The radius of gyration of PEO was approximately $73$ nm, 10-100$\times$ smaller than the contour lengths of WLM chains \cite{armstrong2004hydrodynamic, flory1953principles}
We observed little-to-no reversal in the angular displacement (see blue data in Fig.~\ref{fig:Fig4}B), yielding similar results to those of our control experiments conducted in water only. 
Given that short chains did not exhibit strong reversal, we hypothesized that the long chains of the WLMs and their long-distance entanglements are critical for the large and slow flow reversals. 
Lastly, focusing on WLM solution in MQ water, we observed that increasing the number of windings resulted in greater amount of reversal (Fig.~\ref{fig:Fig4}C). 

Direct numerical solutions of Eqn.~\ref{eq:cauchy} and Eqn.~\ref{eq:continuity} using the upper-convected Maxwell-Oldroyd model also yield reversal (see Fig.~\textcolor{blue}{S9} in \textcolor{red}{SI Appendix}). 
However, it is very weak, $(\theta - \theta_{\text{stop}}) \sim \mathcal{O}(10^{-3})$, and saturates within the relaxation time of the fluid, $(t-t_{\text{stop}}) \sim \mathcal{O}(\lambda)$. 
In the experiments, $(t-t_{\text{stop}}) \sim \mathcal{O}(10\lambda)$. 
The upper-convected Maxwell-Oldroyd model computes the polymeric stress exerted by short, non-interacting dumbbells, and therefore we did not anticipate these results would recapitulate the behavior of WLMs comprised of long chains with significant long-ranged entanglements.

\subsubsection{Bead-spring simulations}
Results from experiments and numerical solutions suggest the presence of an opposing fluid flow to the motion of the driven particle, inducing reversal in the angular displacement of the tracer probe.
To gain further mechanistic insight, we utilize particle-based, non-Brownian Stokesian dynamics (SD) simulations. 
Two torque-free colloidal particles are immersed in a bath of Hookean bead-spring chains with a Newtonian fluid background, modeling a viscoelastic composite medium.

Note that our simple model is heavily coarse-grained and is intended to capture the Hookean elasticity between a large network of entanglements.
Our goal is to capture the weak elasticity generated by many WLM entanglement sites separated by large distances, $\sim \mathcal{O}$(\textmu m), not to model individual WLM chains \cite{boek2002molecular, yakovlev2007molecular, sangwai2011coarse,dhakal2015topology, dhakal2017anomalous}. 
The driven particle generates Newtonian fluid flows that entrain the stationary probe and the monomeric beads (see \textcolor{red}{Movie S4}). 
As a result, bonded beads undergoing relative motion exert elastic forces on the solvent. 
We are interested in the long-ranged coupling of these elastic forces onto the rotational motion of the stationary probe.

We varied the length of bead-spring chains to determine its role in the reversal of the probe's angular displacement.
We present the results of our calculations in Fig.~\ref{fig:Fig4}D after winding the polymeric bath once.
Our results represent an average over many possible realizations of initial bead-spring configurations.
We consider two kinds of chains: short dumbbells whose equilibrium bond length is the monomeric bead diameter, and linear trimers with longer resting bond lengths equal to the trapped colloidal radii (compare inset graphics).
Longer chains reproduced reversal after cessation (compare orange and blue data). 
Further, the amount of reversal increased with increasing number of windings (Fig.~\ref{fig:Fig4}E). 
However, the angular displacement exhibited less reversal from 1.0 to 2.0 windings, compared to the difference from 0.5 to 1.0 winding (compare orange to purple data and green to orange data, respectively).
We do not discern such a trend in the experiments after 30 windings (Fig.~\ref{fig:Fig4}C).

\begin{figure}[t!]
    \centering
    \includegraphics[width=17.8cm]{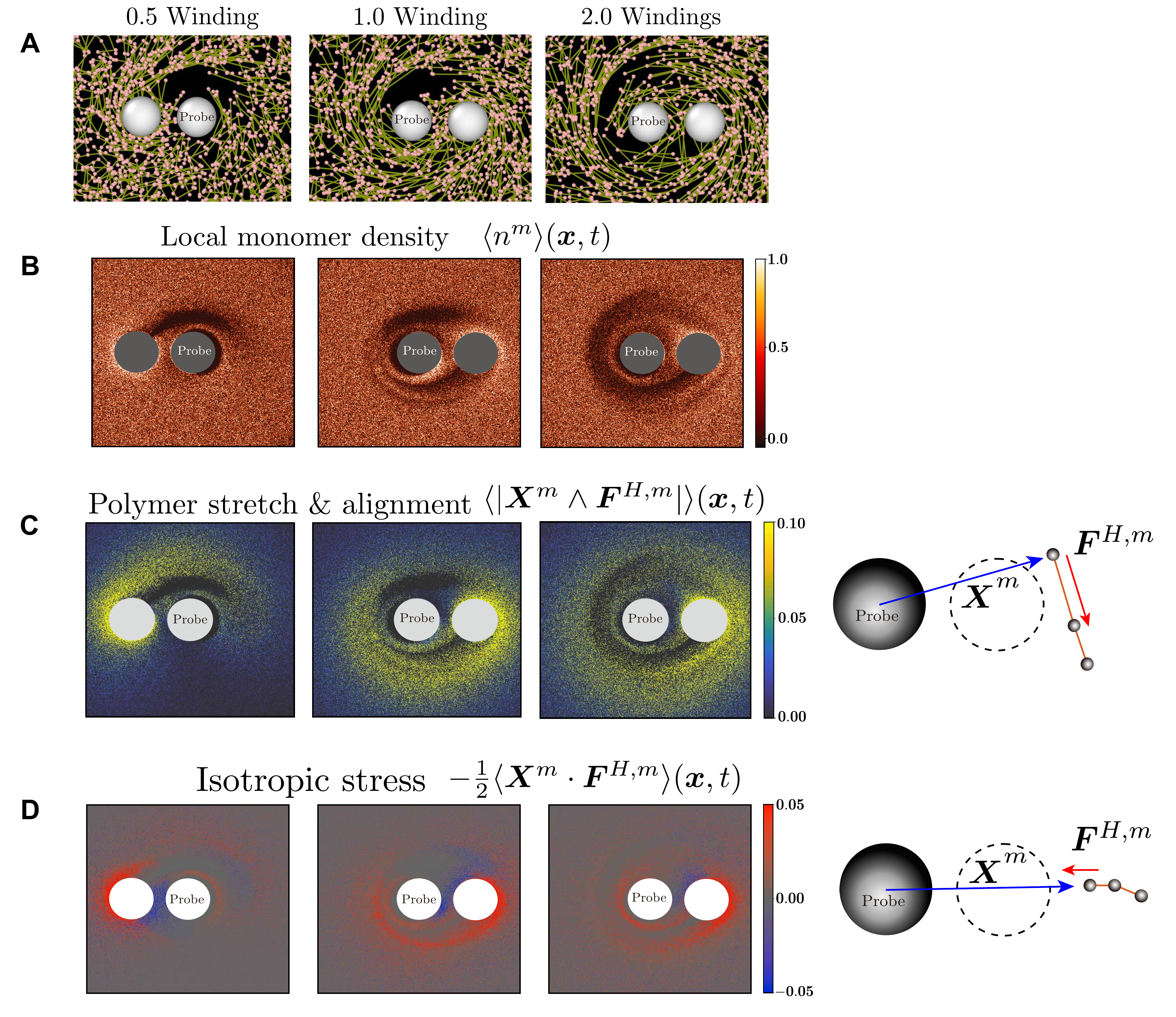}
    \caption{The strain applied by a driven colloidal particle generates stress in a polymeric bath.
We performed non-Brownian Stokesian dynamics simulations of two colloidal particles surrounded by linear bead-spring trimers to model a viscoelastic solution. 
See Materials and Methods for further simulation details.
The center-to-center distance between the two particle is maintained at $r=2.8a_c$, where $a_c$ is the colloidal radius.
(A) From left to right, sample simulation snapshots of trimers after 0.5, and 1.0, and 2.0 windings completed by the moving sphere.
Monomeric beads are shown in pink, and bond connections are highlighted in yellow.
(B) Averaging over many realizations of initial bead-spring configurations, we obtain a measurement of the local monomer density, $\langle n^m \rangle(\bm{x},t)$.
Monomers concentrate at the leading edge of the moving particle and are depleted in the wake (compare white and dark red colors, respectively).
(C) We quantify the bonds' stretch and their alignment by considering the magnitude of the torque from each monomeric bead position and its hydrodynamic force, $\langle |\bm{X}^{m} \wedge \bm{F}^{H,m}|\rangle(\bm{x},t)$.
It is largest on the outward face of the moving sphere (see bright yellow colors from 0.5-2.0 windings).
(D) Simultaneously, part of the hydrodynamic forces on the monomeric beads are directed toward the probe, producing normal stresses that are largest on the outward face of the moving particle and smallest in the intervening gap (compare red and blue colors).
See \textcolor{red}{Movie S5} for animations of panels B-D during winding and relaxation after cessation. 
    }
    \label{fig:Fig5}
\end{figure}

The driven particle distorts the microstructure of the polymeric bath, giving rise to a non-uniform density profile (see \textcolor{red}{Movie S5}).
Monomeric beads accumulate at the leading edge of the moving colloid and deplete in the wake.
Compare the location of monomers in sample snapshots in Fig.~\ref{fig:Fig5}A to brightest and darkest regions in local monomer density, $\langle n^{m}\rangle(\bm{x},t)$, in Fig.~\ref{fig:Fig5}B.
Elastic retraction in the wake of the moving particle homogenizes the local monomer density \cite{khan2019optical}.
Once again, individual bead-spring dumbbells in our simulations are intended to recapitulate the many entanglements in a WLM network, as opposed to single WLM chains.
Therefore, the ``monomeric beads'' should be interpreted as the density of entanglements.

To identify spatially where chains align and experience tensile forces, we computed the magnitude of the hydrodynamic torque, $\langle |\bm{X}^{m} \wedge \bm{F}^{H,m}|\rangle(\bm{x},t)$, generated at each of the beads.
$\bm{X}^{m}$ is the position of a bead relative to the probe, and $\bm{F}^{H,m}$ is its hydrodynamic force (Fig.~\ref{fig:Fig5}C).
As the hydrodynamic forces are balanced by the spring forces, larger values of this metric correspond to greater bond extension or compression. 
From the definition of the cross product, we quantify chain alignment by the degree of perpendicularity between the bond forces and the monomer position (see schematic to far right of  Fig.~\ref{fig:Fig5}C). 
The brightest regions in Fig.~\ref{fig:Fig5}C show that chains are stretched and aligned the most about the driven particle during winding.
Further, chains remain aligned in the wake of the driven particle, whose dimmer colors indicate lasting relaxation of bond forces.
Compare bond connections in simulation snapshots of Fig.~\ref{fig:Fig5}A to corresponding field plots at 0.5, 1.0, and 2.0 windings in \ref{fig:Fig5}C.  

The slow relaxation of restorative elastic forces induces reversal in the angular displacement of the probe. 
The motion of the driven particle generates local fluid flows that both align and stretch chains in its wake (Fig.~\ref{fig:Fig5}C).
Meanwhile, monomers build-up at the leading edge (Fig.~\ref{fig:Fig5}B).
At cessation, the strongest elastic forces on the beads are localized to the leading edge, and bonds are more relaxed in the wake. 
These restorative forces produce fluid disturbances that entrain the probe, which we observe as reversal in its angular displacement.
Put another way, hydrodynamic forces are stored in stretched bonds and slowly dissipate as the polymeric bath relaxes, while the colloidal configuration is held fixed.

The flexibility and length of the chain are critical for the monomeric beads to sample spatial variations in the flow fields generated by the driven particle.
Compared to short dumbbells, this results in larger relative velocities between beads in a bonded pair, enhancing elastic forces that, in turn, produce stronger reversal flows (Fig.~\ref{fig:Fig4}D).
Likewise, winding the polymeric bath continues to stretch the chains, generating stronger elastic forces. 
At the same time, however, this depletes chains about the two colloids (see dilute region in Fig.~\ref{fig:Fig5}B at 2.0 windings).
As the beads are farther away from the probe, their elastic forces produce more distant fluid disturbances that decay over a larger distance. 
Thus, we get saturation in the amount of reversal with greater winding (Fig.~\ref{fig:Fig4}E).

We now reconcile our experimental results with those from our direct numerical solutions and SD simulations.
First, we found little-to-no reversal in our control experiments using PEO as the suspending fluid (Fig.~\ref{fig:Fig4}B).
Likewise, both SD simulations with short dumbbell chains and direct numerical solutions to the upper convected Maxwell-Oldroyd model yielded no reversal (Fig.~\ref{fig:Fig4}D).
These results are consistent with simple Maxwell-Oldroyd constitutive models that assume a uniform spatial distribution of short, non-interacting, harmonic dumbbells.
We obtained significant flow reversals in solutions of WLMs, and we corroborated these results with SD simulations using longer chains.

Based on these insights, our proposed mechanism for flow reversal in the experiments is: (1) the driven colloid stretches the surrounding WLM network, generating an anisotropic, heterogeneous distribution of entanglement sites; 
(2) upon cessation, the network relaxes by loosening long-distance entanglements on slow timescales; 
(3) this structural recovery exerts restorative elastic forces that induce flow reversal and hydrodynamically reverses the probe's angular displacement in the opposite direction. 

Bead-spring models are an oversimplification of WLMs, and our SD model is not intended to recapitulate the detailed structure of WLM chains and their entanglements. 
Rather, our goal is to demonstrate that long, Hookean dumbbells alone are sufficient to reproduce the flow reversals observed in the experiments, which cannot be predicted using simple continuum models like Maxwell-Oldroyd.
Whereas such models assume that the dumbbells stretch under locally linear flow, the discrepancy between the continuum calculations and experiments suggests that entangled WLM networks exhibit highly non-local coupling with colloidal motion.
Therefore, we believe that WLM chain-chain entanglements spanning colloidal length scales are critical for producing the nonlinear creep response of the tracer probe.

We also tested whether flow reversal is observable in a conventional macrorheometer.
We conducted creep and recovery tests in a stress-controlled rheometer, and indeed found that WLM solutions generate flow reversals in the angular displacement of the rheometer plates during recovery, albeit requiring much larger imposed stresses (see Fig.~\textcolor{blue}{S12}A in \textcolor{red}{SI Appendix}).
As a control, we observed no reversal using solutions of small molecule constituents, glycerol and PEO (see Fig.~\textcolor{blue}{S12}B and C), supporting our hypothesis that nonlinear straining of entangled networks, capable of storing long-ranged viscoelasticity, produces flow reversal upon cessation.

\subsubsection{Colloidal attraction}

\begin{figure}[t!]
    \includegraphics[width=\textwidth]{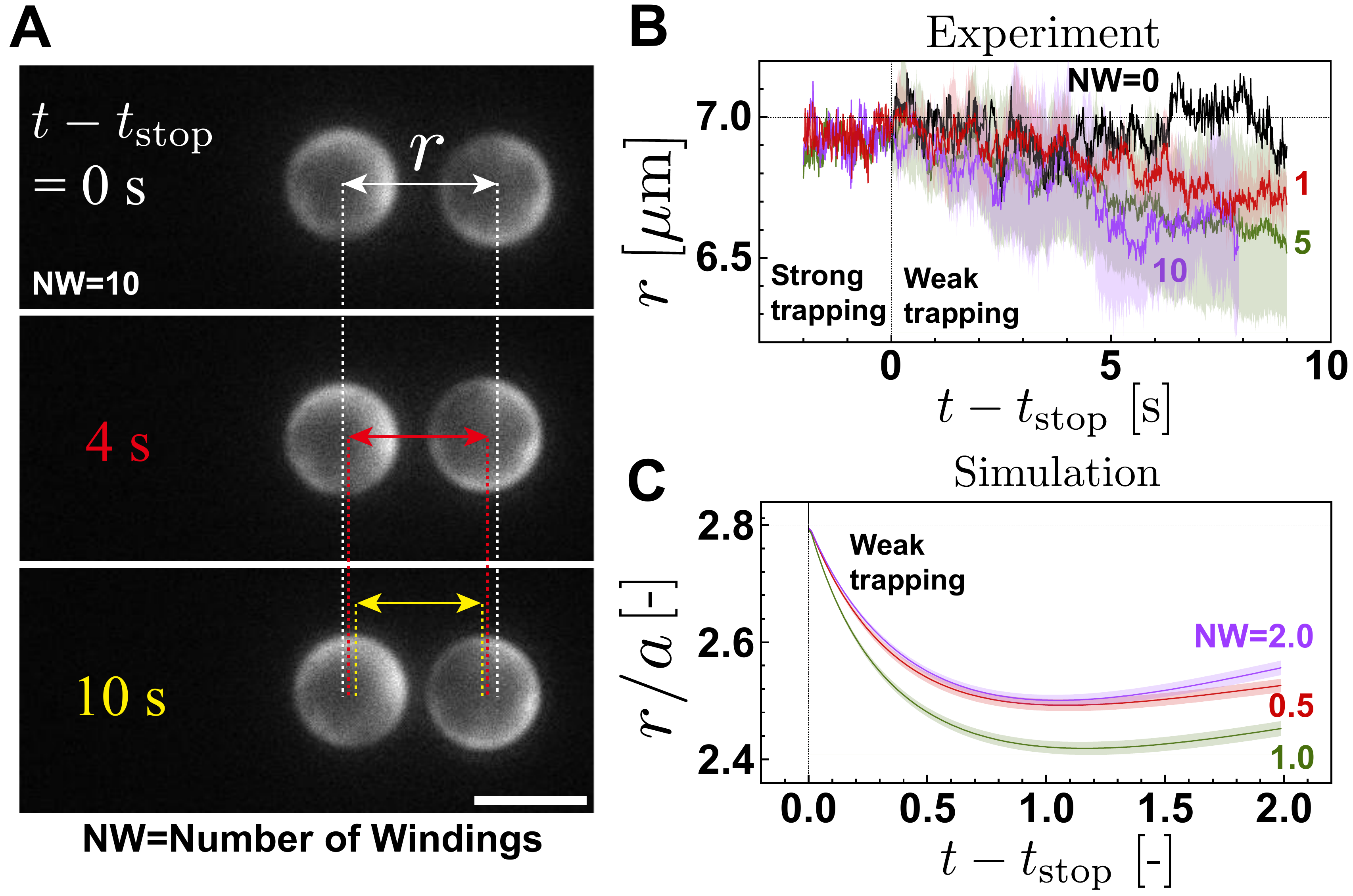}
    \centering
    \caption{Straining wormlike micellar solution with a driven colloidal particle induces a hydrodynamic ``attraction''. We used stiff traps to maintain the center-to-center distance between a pair of spheres immersed in a WLM solution at $r  = 2.8a = 7.0$ \textmu m (``strong trapping''). 
    We drove a particle around a stationary probe in a circular trajectory for a number of periods (``windings''), stopping at $t_{\text{stop}}$. (A) Timelapse shows shrinking center-to-center distance, $r$, between two colloidal particles upon lowering the trap stiffness. Scale bar is 5~\textmu m. See \textcolor{red}{Movie S6} for a video of the experiment. 
    (B) Evolution of center-to-center separation after 1, 5, and 10 windings (``weak trapping'').  
    (C) We corroborated our observations with non-Brownian Stokesian dynamics simulations of a pair of spheres in a bath of linear bead-spring chains (see Fig.~\ref{fig:Fig4}).
    After 0.5, 1.0, and 2.0 windings, we reduced the trapping stiffness on the spheres and tracked the evolution of their center-to-center distance (green, orange, and purple data, respectively). 
    Here, time is scaled by the period of orbit of the moving particle. 
    See \textcolor{red}{Movie S7} for a sample simulation video.
    }
    \label{fig:Fig6}
\end{figure}

In our simulations, we observed positive normal stresses developing on the outward faces of the moving particle and probe, while an opposing stress develops in their interfacial gap. 
To these ends, we computed the isotropic stress exerted by the chains, $-\frac{1}{2}\langle\bm{X}^{m} \cdot \bm{F}^{H,m}\rangle(\bm{x},t)$ in Fig.~\ref{fig:Fig5}D (see also schematic to the far right).
Simultaneously, monomers are enriched in the bulk region outside of the two colloids, and polymeric stresses on the outward faces are greater than those in the interfacial gap (compare Fig.~\ref{fig:Fig5}B to Fig.~\ref{fig:Fig5}D).
This imbalance of stress generates fluid flows that would bring the colloids together, presenting as a hydrodynamic ``attraction'' reminiscent of depletion flocculation.
Direct numerical solutions of Eqs.~\ref{eq:cauchy} and \ref{eq:continuity} using the upper-convected Maxwell-Oldroyd model show that the colloids experience decaying attraction forces, following cessation (see Fig.~\textcolor{blue}{S10} in \textcolor{red}{SI Appendix}).
Previous studies report colloidal flocculation in simulations and experiments of particle suspensions with viscoelastic suspending media under shear \cite{Hu2022-js, Santos-de-Oliveira2013-hx}.

Inspired by the simulations, we returned to experiments and applied a step-decrease in the laser stiffness on the trapped colloids after cessation. 
Indeed, the two particles fluxed towards each together, as their center-to-center separation decreased with time (see timelapse in Fig.~\ref{fig:Fig6}A and \ref{fig:Fig6}B, respectively).
See \textcolor{red}{Movie S6} for a video of the experiment.
At the end of the observation window, the separation distance between the two particles became smaller with increasing number of windings (compare red, green, and purple data in Fig.~\ref{fig:Fig6}B). 
We also present the results of separate SD simulations in Fig.~\ref{fig:Fig6}C, where we also lowered the trapping stiffness on the colloids upon cessation.
See \textcolor{red}{Movie S7} for a simulation video.
At later times in the simulations, the relaxation of the chains reduces their overall force on the particles, and the center-to-center distance between the colloids increases as they are pulled toward their respective trap positions (see increasing separation distance at times $t \gtrsim 1.0$ in Fig.~\ref{fig:Fig6}C).

We further note that, in our simulations, the minimum center-to-center distance between the two colloids maximizes at 1.0 winding and increases after 2.0 windings.
Similar to the diminishing amount of reversal in Fig.~\ref{fig:Fig4}E with increasing number of windings, the depletion of monomeric beads about the colloids results in the transmission of elastic forces that propagate over larger separations.
Nevertheless, our simulations demonstrate that long-ranged fluid disturbances couple elastic forces in a polymeric medium to the dynamics of colloids, manifesting as time-dependent HIs. 

\subsection{Many-body interactions} 
Until now, we have focused on the pair hydrodynamic interaction between two colloidal particles.
We now consider more complex multiparticle configurations, where many-body hydrodynamic interactions significantly influence the mobility of a given particle.
For simplicity, we focus on the linear viscoelastic response of the suspending medium.
We expand on our two-body microhydrodynamic analysis from earlier and arrive at a saddle-point formulation of  Stokesian dynamics \cite{Fiore2019-or} that holds for linear viscoelastic suspending media. 

\begin{figure}[b!]
\centering
\includegraphics[width=\textwidth]{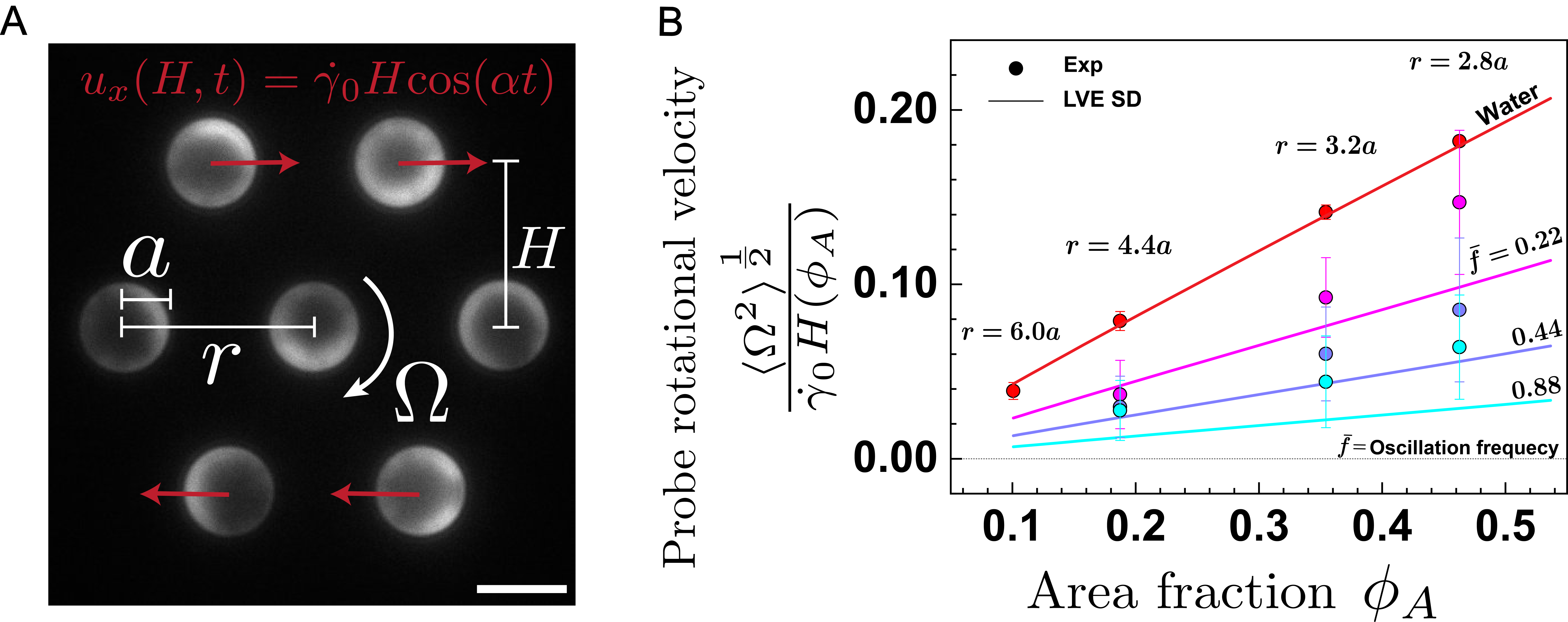}
\caption{
Nearest neighbors hinder rotational mobility during shear of a two-dimensional hexagonal colloidal
cluster in a viscoelastic fluid. (A) In a WLM solution with relaxation time $\lambda$, we trapped seven particles of radius $a = 2.5$ \textmu m into a configuration that mimicked the first coordination layer of a hexagonal lattice with pairwise distances $r = 2.8a - 6.0a$. These corresponded to local area fractions $\phi_A = 2\pi a^2(r^{2}\sqrt3)^{-1} \approx 10\% - 46\% $. To model an oscillatory shear strain on the lattice, we imposed a velocity profile in the x-direction of the form $u_x(y,t) = \skew{4}{\dot}{\gamma}_0y\cos(\alpha t)$, with shear-rate amplitude $\skew{4}{\dot}{\gamma}_0 $ and oscillation frequency $\alpha$. $H(\phi_A) = r(\phi_A)\cos(\pi/6)$ is half of the total height of the structure. 
See \textcolor{red}{Movie S8} for a video of the experiment. 
The Weissenberg numbers are $\text{Wi}=\skew{3}{\dot}{\gamma}_0\lambda\sim \mathcal{O}(10^{-2} - 10^{-1})$. 
Scale bar is 5 \textmu m. 
We define a dimensionless frequency, $\bar{f} = \lambda \alpha/2\pi$, comparing the relaxation time of the WLM solution to the linear frequency of the straining motion.
At a given area fraction and oscillation frequency, we measured the time-dependent rotational velocity of the center sphere, $\Omega$.
(B) Time-averaged magnitudes of the center sphere's rotational velocity are plotted as a function of the area fraction across different values of $\bar{f}$ (colored points).  
Corresponding results in water are shown for comparison (red points).
To corroborate our measurements, we present results from Stokesian dynamics calculations in a linear viscoelastic medium, using Eqn.~\ref{eq:lve_saddle_point} ("LVE SD", solid lines).
}
\label{fig:Fig7}
\end{figure}

In our previous study \cite{kimnagella2025direct}, we trapped seven particles in a configuration mimicking the first coordination layer of a hexagonal lattice (see Fig.~\ref{fig:Fig7}A). 
All particles were equidistant from one another with spacing, $r$, that was related to the local area fraction, $\phi_A = 2\pi a^2(r^2\sqrt 3)^{-1}$. 
Our aim was to measure the rotational mobility of the center sphere as a function of the packing parameter, $\phi_A$, in water.
We imposed an oscillatory strain on the structure, where the top and bottom rows of spheres translated in opposite directions along the x-direction according to $u_x(y,t) = \skew{4}{\dot}\gamma_0y\cos(\alpha t)$, and we measured the rotational velocity of the center probe as a function of time.
We now perform these experiments in WLM solution, applying small-amplitude oscillatory strain.

Having multiple nearby particles generates additional resistance to the flow of fluid disturbances due to the particles' impermeable boundaries.
Accounting for the induced force-dipoles, or stresslets, approximates the influence of many-body HIs.
We write the following frequency-dependent mobility relation,
\begin{equation}\label{eq:lve_mobility}
     \mathcal{M}^{\dagger}(\bm{r}^{N})\cdot \hat{\mathcal{F}}(\omega) = \hat{\mathcal{U}}(\omega)\hat\varphi^{-1}(\omega),
\end{equation}
where $\mathcal{M}^{\dagger}(\bm{r}^{N}) = \mathcal{M} - \textbf{M}^{\mathcal{U}\text{S}} \cdot \left( \textbf{M}^{\text{ES}} \right)^{-1} \cdot \textbf{M}^{\text{E}\mathcal{F}}$ is the grand mobility matrix with stresslet constraints \cite{Durlofsky1987-oc, Fiore2018-ax, kimnagella2025direct}.

We seek to predict the induced rotational motion of the center sphere in Fig.~\ref{fig:Fig7}A.
We couple Eqn.~\ref{eq:lve_mobility} to the torque balance, $\skew{-4}{\hat}{\mathbf{L}}^{\text{H}} = \bm{0}$.
After some manipulation and reverting to the time domain, we arrive at the saddle-point problem,

\begin{equation}\label{eq:lve_saddle_point}
\begin{bmatrix}
        \mathcal{M}^{\dagger} & \mathcal{B} \\
        \mathcal{B}^{T} & \bm{0}
    \end{bmatrix}
    \begin{bmatrix}
        \mathcal{F}(t) \\
        \mathcal{Y}(t)
    \end{bmatrix}
    =
    \begin{bmatrix}
            \mathcal{W}(t) \\
         \bm{0}
\end{bmatrix},
\end{equation}
whose solution gives the desired angular velocities, $\mathcal{Y}(t) = \int_{-\infty}^t \bm{\Omega}(s)\varphi^{-1}(t-s) \text{d}s$, forced by $\mathcal{W}(t) = \left[ \int_{-\infty}^t \mathbf{U}(s)\varphi^{-1}(t-s) \text{d}s  \quad \bm{0} \right]^{T}$.
The projection operator, $\mathcal{B}$, is defined so that  $\left[\hat{\mathbf{U}} \quad \bm{0} \right]^{T}-\mathcal{B} \cdot \hat{\bm{\Omega}} = \hat{\mathcal{U}}$. 

We apply the displacement on the top and bottom rows of spheres in the WLM solution and compute the time-averaged rotational mobility of the center sphere,
$\langle \Omega^2 \rangle^{1/2}(\phi_A) = \\ \left( \frac{1}{T}\int_{t}^{t+T}\Omega^2(t'; \phi_A)\text{d}t' \right)^{1/2}$, as a function of the area fraction. 
The driving force for fluid disturbances in Eqn.~\ref{eq:lve_saddle_point} is

\begin{equation} \label{eq:Maxwell_driving_force}
\begin{split}
    \int_{-\infty}^t u_{x}(y,s)\varphi^{-1}(t-s)\text{d}s =
    \skew{3}{\dot}{\gamma}_0y\frac{2\pi\bar{f} \sin({\alpha t}) + \cos(\alpha t)}{1 + \left(2\pi\bar{f} \right)^2} ,
\end{split}
\end{equation}
where $\bar{f} = \lambda \alpha/2\pi$ is a dimensionless linear frequency.

Figure \ref{fig:Fig7}B presents the results of the center sphere's time-averaged rotational mobility from both experiments and simulations (compare points and solid lines, respectively).
The zero-frequency limit, $\bar{f} = 0$, corresponds to a Newtonian fluid (see red data).
As we increase the oscillation frequency the medium exhibits a delayed response to the imposed driving forces. 
Thus, the center sphere in the hexagonal structure is entrained in a hindered flow (see purple data in for $\bar{f}=0.22$).
This attenuation of the disturbance flow is driven by $\mathcal{W}(t) \sim \mathcal{O}\left(\bar{f}^{-1}\right)$ in Eqn.~\ref{eq:Maxwell_driving_force}. 
Therefore, the mobility of the center sphere montonically decreases with the oscillation frequency (see data for $\bar{f} = 0.44$ and $0.88$).

\section{Discussion and Conclusion} \label{sec:conclusion}
In this work, we showed that ``time-dependent'' colloidal HIs in viscoelastic fluids arise from structural memory generated in the surrounding medium.

Using solutions of wormlike micelles as a model viscoelastic fluid, we measured the induced angular displacement of a stationary probe due to the circular motion of a nearby driven particle.
We obtained a distribution of relaxation times in WLM solution from transient start-up, suggesting non-negligible spatial variation in entanglement density on colloidal length-scales. 
After sudden cessation of the driven particle, the angular displacement of the probe reversed for several seconds.
Direct numerical solutions predicted reversal flows after cessation, but were much weaker and shorter-lasting than those inferred from the experiments.
These discrepancies reveal a breakdown of the continuum model for the viscoelastic stress. 
Specifically, Maxwell-Oldroyd-type models are developed from coarse-graining the dynamics of dumbbells whose ends sample locally linear flow, assuming the length-scale of the flow to be much greater than that of the dumbbells.
However, our Stokesian dynamics calculations showed that longer chains, with resting lengths comparable to the size of colloids, reproduced reversal after cessation.
Therefore, we believe that entangled networks of WLMs exhibit highly non-local coupling with colloidal motion to produce the striking and persistent signals from the tracer probe.  

One distinguishing feature of our optical tweezer setup is the ability to investigate mesoscale heterogeneities that are inaccessible using conventional macrorheometry.
The shear deformation imposed by our moving colloid is highly localized at the WLM lengths and network entanglement sizes, and thus the fluid disturbances reported by the tracer probe are sensitive to local structural heterogeneites.
These heterogeneities produce the observed variance throughout our linear and nonlinear viscoelastic measurements when we adjust the placement of the two spheres.
In contrast, a macrorheometer applies uniform deformation across the bulk sample volume, so that all spatial variations are averaged out. 
Further, the required applied stress to observe appreciable flow reversal in the macrorheometer was $\mathcal{O}(100\,\text{Pa})$, compared to $\mathcal{O}(1\,\text{Pa})$ with optical tweezers.

Another major and fundamental difference of our setup from conventional macrorheometry is the geometry.
Because the colloids are fully immersed within the complex media, our two-particle flow geometry is designed to measure hydrodynamic interactions between colloidal probes.
In microrheology, a single probe is inserted into the medium to characterize the fluid viscoelasticity.
The probe is typically chosen to be larger than the length scales of the constituents and entanglements to ensure a macroscopic measurement of transport coefficients. 
Our two-particle setup is designed to directly quantify interactions among colloids, including hydrodynamic and osmotic depletion-like phenomena produced by driving the suspending medium out of equilibrium.

We conclude by suggesting areas of further investigation.
We introduced a simulation method, based on Eqn.~\ref{eq:lve_saddle_point}, to compute colloidal dynamics due to many-body HIs in linear viscoelastic media.
Our formulation may be readily implemented in existing large-scale simulation codes \cite{Fiore2019-or} to study the rheology of non-Brownian colloidal dispersions in viscoelastic suspending media.
We considered materials whose linear viscoelasticity exhibits unimodal stress relaxation, like WLM solutions, but the present formulation can be extended to an arbitrary number of modes.
Incorporating thermal forces in our simulation method is an ongoing effort. 
We can include $\mathcal{O}(\text{Wi})$ effects from the upper-convected Oldroyd model in our SD simulations.
Phillips has proposed similar modifications to the Stokesian dynamics formalism to account for nonlinear viscoelasticity \cite{Phillips1996-fy, Phillips2003-nj}.
Further, while we focused on solutions of wormlike micelles, our experiments may be performed in any number of complex fluid media to probe length scales that cannot be probed by traditional macroscopic rheometers.

\section*{Acknowledgments}
This material is based upon work supported by the National Science Foundation under Grant No.~2440029.
S.C.T. is also supported by the Packard Fellowship in Science and Engineering.
Some of the computing for this project was performed on the Sherlock cluster. We would like to thank Stanford University and the Stanford Research Computing Center for providing computational resources and support that contributed to these research results.

\bibliographystyle{unsrt}
\bibliography{pepadun}

\end{document}